\documentclass[aps,prl,twocolumn,superscriptaddress]{revtex4}
\usepackage{amsmath}
\usepackage{bm}
\usepackage{graphicx}
\usepackage{times}
\usepackage{color} %for change tracking
\newcommand{\ket}[1]{|#1\rangle}

\newcommand{\eq}{\begin{equation}}
\newcommand{\fine}{\end{equation}}

\begin{document}
\title{Complete experimental toolbox for alignment-free quantum communication}

\author{Vincenzo D'Ambrosio}
\author{Eleonora Nagali}
\affiliation{Dipartimento di Fisica, Sapienza Universit\`{a} di Roma, Roma 00185, Italy}
\author{Stephen P. Walborn}
\affiliation{Instituto de F\'isica, Universidade Federal do Rio de Janeiro, Rio de Janeiro, RJ 21941-972, Brazil}
\author{Leandro Aolita}
\affiliation{ICFO-Institut de Ci\`{e}ncies Fot\`{o}niques, Av. Carl Friedrich Gauss 3, 08860 Castelldefels (Barcelona), Spain}
\author{Sergei Slussarenko}
\affiliation{Dipartimento di Scienze Fisiche, Universit\`{a} di Napoli ``Federico II'',
Compl.\ Univ.\ di Monte S. Angelo, 80126 Napoli, Italy}
\author{Lorenzo Marrucci}
\affiliation{Dipartimento di Scienze Fisiche, Universit\`{a} di Napoli ``Federico II'',
Compl.\ Univ.\ di Monte S. Angelo, 80126 Napoli, Italy}
\affiliation{CNR-SPIN, Complesso Universitario di Monte S. Angelo, 80126 Napoli, Italy}
\author{Fabio Sciarrino}
\email{fabio.sciarrino@uniroma1.it}
\affiliation{Dipartimento di Fisica, Sapienza Universit\`{a} di Roma, Roma 00185, Italy}

\begin{abstract}

\end{abstract}
\maketitle
%%%%%%%%%%%%%%%%%%%%%%%%%%%%%%%%%%%%%%%%%%%%%%%%%%%%%%%%%%%%%%%%%%%%%%%%%%
\textbf{Quantum communication employs the counter-intuitive features of quantum physics to perform tasks that are impossible in the classical world. It is crucial for testing the foundations of quantum theory and promises to revolutionize our information  and communication technologies. However, for two or more parties to execute even the simplest quantum transmission, they must establish, and maintain, a shared reference frame. This introduces a considerable overhead in communication resources, particularly if the parties are in motion or rotating relative to each other.  We experimentally demonstrate how to circumvent this problem with the efficient transmission of quantum information encoded in rotationally invariant states of single photons. By developing a complete toolbox for the efficient encoding and decoding of quantum information in such photonic qubits, we demonstrate the feasibility of alignment-free quantum key-distribution, and perform a proof-of-principle alignment-free entanglement distribution  and violation of a Bell inequality. Our scheme should find applications in fundamental tests of quantum mechanics and satellite-based quantum communication.}

\begin{figure}
\begin{center}
\includegraphics[width=8cm]{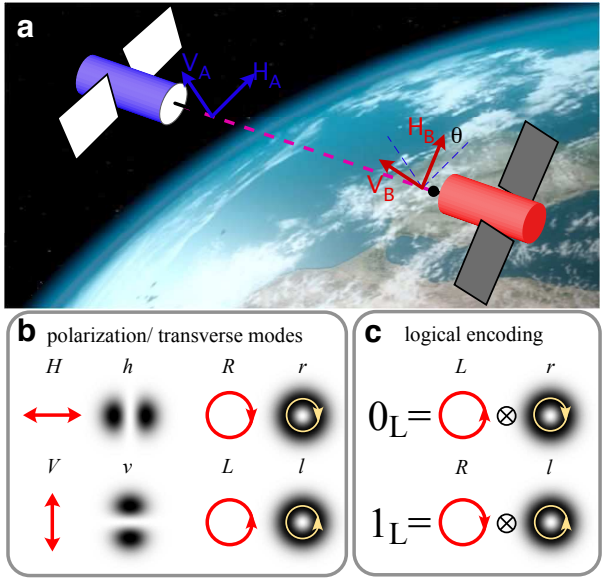}
\caption{{\bf Misalignment-immune single-photon qubits.} \textbf{a)} Alice and Bob, here depicted as satellites, need to carefully control the relative orientation between their horizontal ($H$) and vertical ($V$) axes to faithfully implement quantum communication in free-space. Unknown misalignments around the propagation axis manifest as rotations of the transmitted qubits by unknown angles $\theta$ in the $H-V$ plane.  \textbf{b)} Qubits can be equivalently encoded in both polarization and transverse modes: $H/V$ denote horizontal/vertical linear
polarizations, $L/R$ left/right circular polarizations, $h/v$
horizontal/vertical first-order Hermite-Gauss modes, and $l/r$ left- and
right-handed first-order Laguerre-Gauss modes. The $L/R$ polarizations are eigenstates with
eigenvalues $\pm\hbar$ of the spin angular momentum (SAM), whereas the $l/r$ modes are the equivalent eigenstates of the orbital angular momentum (OAM). \textbf{c)} By combining SAM and OAM eigenstates of opposite handedness, two null-eigenvalue eigenstates of the total angular momentum arise. Both these hybrid states are invariant under rotations around the propagation axis, and can therefore encode misalignment-immune logical qubit states, called $0_{\text{L}}$ and $1_{\text{L}}$.}
 \label{fig:1}
\end{center}
\end{figure}

All current implementations of quantum communication (QC) use
photons as the carriers of qubits (quantum bits), the basic units of quantum information. This is due to the fact that photons are easy to transport from one location to another, known as so-called ``flying qubits'' \cite{Gisi02}. Photonic free-space QC has been demonstrated for distances of hundreds of kilometers \cite{Ursi07}, a progress that could soon
lead to satellite-based long-distance QC \cite{Rari02,Aspe03,Vill08,Bona09}. However,
standard approaches to QC, for example based on encoding qubits into the polarization of photons, require that all users involved have knowledge of a \textit{shared reference frame} (SRF). For instance, in the bipartite scenario, the emitter and receiver, conventionally called Alice and Bob, must initially align their local horizontal ($H$) and vertical ($V$) transverse axes, and then keep them aligned throughout the transmission (see Fig. \ref{fig:1} \textbf{a}). This in turn requires the exchange of a large (strictly speaking, infinite) amount of classical information. This represents, in general, an extra technical overhead, which can impose very serious obstacles in the particular situations where the users are very far apart from each other, the misalignment between their frames varies in time, or the number of users is large, for example \cite{Pere01,Bart07}. In general, the lack of a SRF inhibits faithful QC because it is equivalent to an
unknown relative rotation, therefore introducing noise into the quantum
channel \cite{Bart07}. 

A possible solution to this problem is to exploit multi-qubit entangled states that are invariant under single-qubit rotations acting collectively on all the qubits (see \cite{Cabe03,Walt03,Boil04,Bour04,Chen06} and references therein). These constitute particular instances of decoherence-free subspaces, originally introduced  in the context of fault-tolerant quantum computing  \cite{Palm96, Bare97,Zana97,Lidar98}. The idea is thus to encode logical qubits into rotationally invariant states of multiple physical qubits. These can in principle be realized with multiple photons \cite{Walt03,Boil04,Bour04}. However, the efficient production and detection of multi-photon states is a technological challenge, they are more susceptible to losses, and the requirement that multiple photons are subject to exactly the same rotation is very seldom perfectly satisfied. 

A more efficient way to circumvent misalignments is provided by exploiting multiple
degrees of freedom of single photons \cite{Aoli07}. In particular, the polarization and transverse spatial modes stand out for this purpose (see Fig. \ref{fig:1} \textbf{b}). Just as the circular polarization states are eigenstates of the spin angular momentum (SAM) of light, the helical-wavefront Laguerre-Gaussian (LG) modes  are eigenmodes of its \textit{orbital angular momentum} (OAM). The OAM degree of freedom is attracting a growing interest for applications in both classical and quantum photonics \cite{Moli08,Fran08,Stut07,Marr11}.
The peculiarity about the SAM and (first order)
OAM eigenstates together is that, since they are defined with respect to the same
reference frame, they suffer exactly the same transformation under coordinate rotation. Therefore, they satisfy the collective rotation requirement exactly, constituting an ideal pair to carry rotationally invariant hybrid qubits (see Fig.\ \ref{fig:1}\textbf{c}). 

\begin{figure*}
\includegraphics[scale=.45]{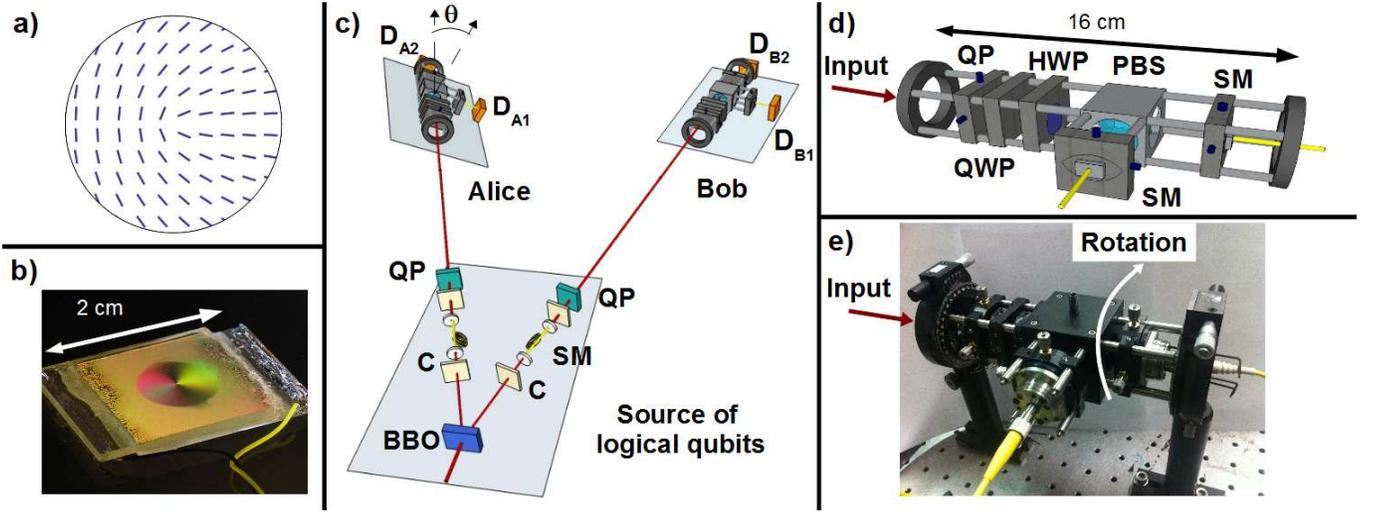}
\caption{{\bf Toolbox for experimental alignment-free quantum communication.} \textbf{a-b)} The liquid crystal QP with topological
charge $q=1/2$ works as a universal logical-qubit encoder/decoder. Panel \textbf{a)} shows the QP optical axis pattern, while
panel \textbf{b)} is a photo of the device seen through crossed polarizers, under oblique illumination; different colors result from different
optical axis orientations. The QP birefringent retardation is electrically tuned. \textbf{c)} Experimental setup, in the
configuration used to generate entangled rotationally invariant photon pairs and perform a  misalignment-immune demonstration of non-locality. Reference-frame misalignments are implemented by physically rotating Alice's entire measurement station around the optical axis by an angle $\theta$. For the alignment-free BB84 QKD test, the entangled-photon source together with Bob's measurement station is taken as the transmitting party, and Bob's photon is used to herald the transmission of the other photon to Alice. The communication distance was 60 cm. \textbf{d)} Schematics of the rotating device for measuring rotationally invariant qubits in arbitrary reference frames. \textbf{e)}
Photo of the actual measurement device. Legend: QP - $q$-plate; BBO - $\beta$-barium borate crystal; C walk-off compensation crystals; SM
- single-mode fibers; D - single photon detectors; HWP - half-wave plate; QWP - quarter-wave plate; PBS - polarizing beam-splitter.}
\label{fig:2}
\end{figure*}

Here we experimentally demonstrate a complete toolbox for the efficient encoding and decoding of quantum
information in  such photonic qubits, suitable for alignment-free QC. The core of our toolbox is a liquid crystal device, named ``$q$-plate'' (see Fig.\ \ref{fig:2} \textbf{a, b}) \cite{Marr06, Marr11}, that maps polarization-encoded qubits into qubits encoded in hybrid polarization-OAM states of the same photon that are invariant under arbitrary rotations around the propagation direction, and
vice versa. In other words, the q-plate acts as a universal
encoder/decoder, where ``universal" refers to the fact that it works for any
qubit state. The $q$-plate used in the present work is the result of a
very recent technological advance allowing for the manufacture of
electrically tunable devices with topological charge $q=1/2$
\cite{Slus11}. This is the first time such devices are exploited in the quantum regime. In addition, the toolbox requires no interferometric stability as in previous proposals \cite{Aoli07,Souz08}, and can be set entirely in a robust and compact unit that could easily be mounted in a small satellite, for instance. Furthermore, our universal-decoder setup features a built-in filtering mechanism that maps a wide class of physical errors into loses instead of logical errors. We show that, due to this mechanism, the scheme is robust also against misalignments around axes other than the propagation direction too, as well as other spatial perturbations. We demonstrate the potential of our method by
performing a proof-of-principle misalignment-immune implementation of the single-photon Bennett-Brassard (BB84) quantum
key-distribution (QKD) protocol \cite{Benn84}, entanglement distribution, 
and the violation of the Clauser-Horne-Shimony-Holt (CHSH) Bell inequality
\cite{Clau69}.

Apart from the studies with multiple-photon states \cite{Cabe03,Walt03,Boil04,Bour04,Chen06}, the interest in misalignment-free QC is also reflected by more recent proposals \cite{Laing10,Lian10,Wall11} and experiments \cite{Shad11}. The latter provide interesting single-photon methods for QKD \cite{Laing10} and non-locality tests \cite{Lian10, Shad11,Wall11}. However, they do not offer protection against general (for instance arbitrary time-varying) misalignments; and, furthermore, they do not allow for generic QC, e.g. for the transmission of arbitrary qubit states.
%%%%%%%%%%%%%%%%%%%%%%%%%%%%%%%%%%%%%%%%%%%%%%%%%%%%%%%%%%%%%%%%%%%%

{\bf Hybrid logical qubit encoding.} Our logical qubit basis is defined by the hybrid polarization-OAM single-photon states 
\begin{eqnarray}
\ket{1}_{\text{L}} &=& \ket{R}_{\text{p}}\ket{l}_{\text{o}}\nonumber\\
\ket{0}_{\text{L}} &=& \ket{L}_{\text{p}}\ket{r}_{\text{o}}.
\label{logic}
\end{eqnarray}
The symbols inside the kets are defined in Fig.\ \ref{fig:1}. Subscript $\text{p}$ denotes the polarization Hilbert space, spanned by the eigenstates $\ket{L}_{\text{p}}$ and $\ket{R}_{\text{p}}$ of the SAM operator $S^z_{\text{p}}$ along the propagation direction ($z$ axis), with respective eigenvalues $s^z_{\text{p}}=\hbar$ and $s^z_{\text{p}}=-\hbar$. In turn, subscript o stands for the OAM bidimensional subspace spanned by the eigenstates $\ket{l}_{\text{o}}$ and $\ket{r}_{\text{o}}$ of the OAM operator $S^z_{\text{o}}$, with respective eigenvalues $s^z_{\text{o}}=\hbar$ and $s^z_{\text{o}}=-\hbar$. Logical hybrid states \eqref{logic} possess zero total angular momentum along $z$, that is, they 
are null-eigenvalue eigenstates of the total SAM + OAM operator $S^z_{\text{p}}+S^z_{\text{o}}$. Therefore, since the total angular momentum operator is the generator of state rotations, states \eqref{logic} are both invariant under arbitrary rotations around the $z$ axis. More specifically, in a physical rotation about the $z$ axis by any angle $\theta$, the circular polarization states and OAM eigenmodes acquire equivalent phase factors on their own: $\ket{L/R}_{\text{p}}\to e^{\mp i\theta}\ket{L/R}_{\text{p}}$ and $\ket{l/r}_{\text{o}}\to e^{\mp i\theta}\ket{l/r}_{\text{o}}$. However, for tensor-product combinations with opposite handedness as \eqref{logic}, the individual phases cancel each other and the composite states  remain intact. As a consequence, because of linearity, any  coherent superposition (or incoherent mixture) of these two logical states is also immune to all possible reference-frame misalignments during the entire QC session. 
%%%%%%%%%%%%%%%%%%%%%%%%%%%%%%%%%%%%%%%%%%%%%%%%%%%%%%%%%%%%%%%%%%%%%%%%%%
\par{\bf Universal encoder/decoder}.
The experimental setup used to encode and decode arbitrary hybrid qubit states in the logical basis  (\ref{logic}) is shown in Fig.\
\ref{fig:2}. The $q$-plates (QPs) are liquid
crystal devices that produce a spin-orbit coupling of the polarization
and OAM contributions to the total angular momentum of photons
\cite{Marr06,Marr11}. The QP is a birefringent slab having a uniform
optical retardation $\delta$ and a suitably patterned transverse
optical axis, with a topological singularity of charge $q$ at its
center. A ``tuned'' QP with $\delta=\pi$ transfers quanta of angular momentum between the SAM and the OAM. Specifically, each photon suffers a variation in its OAM by an amount $\Delta =s^z_{\text{p}} 2q$ determined by the charge $q$ and the SAM $s^z_{\text{p}}$ of the input polarization. QPs with $q=1$ have been recently used to demonstrate interesting spin-OAM quantum information manipulations \cite{Naga09a,Naga09d,Naga10b}.

\par A tuned QP with topological charge $q=1/2$ \cite{Slus11} gives rise to the following transformations:
\begin{eqnarray}
\label{transformations}
\ket{R}_{\text{p}}\ket{0}_{{\text{o}}}\stackrel{QP}{\rightarrow}\ket{L}_{\text{p}}\ket{r}_{\text{o}} &=& \ket{0}_{\text{L}}\nonumber\\
\ket{L}_{\text{p}}\ket{0}_{{\text{o}}}\stackrel{QP}{\rightarrow}\ket{R}_{\text{p}}\ket{l}_{\text{o}}
&=& \ket{1}_{\text{L}}, \label{qplate}
\end{eqnarray}
where $\ket{0}_{{\text{o}}}$ denotes a zero OAM state, such as the fundamental Gaussian mode (TEM$_{00}$). We note that the radial profile of the $\ket{l}_{\text{o}}$ and $\ket{r}_{\text{o}}$ states generated by the $q$-plate is not exactly Laguerre-Gauss (see Supplementary Information for more details), but this does not affect their pure OAM-eigenstate rotational behavior \cite{Marr11}. Indeed the QP is ideally a unitary device (our QP was uncoated and had a transmission efficiency of about $85\%$), but the induced radial-mode effects (see Supplementary Information) may introduce about $40\%$ of total additional losses in the final recoupling to the single-mode fiber before detection. Consider then a generic
polarization-encoded qubit
$\ket{\psi}_{\text{p}}=\alpha\ket{R}_{\text{p}}+\beta\ket{L}_{\text{p}}$ prepared in
the TEM$_{00}$ spatial mode. From transformations \eqref{transformations}, sending  the qubit through the QP yields
\begin{equation}
\ket{\psi}_{\text{p}}\ket{0}_{{\text{o}}} \stackrel{QP}{\rightarrow}
\alpha\ket{0}_{\text{L}} + \beta \ket{1}_{\text{L}} = \ket{\psi}_{\text{L}}. 
\label{generation}
\end{equation}
That is,  the qubit is now encoded into the desired rotationally invariant space
spanned by logical basis \eqref{logic}. Remarkably, the same QP device works also as a universal decoder, transferring
generic rotationally invariant  qubits to their polarization-encoded counterparts. Explicitly, by injecting $\ket{\psi}_{\text{L}}$ into the QP, one obtains
\begin{equation}
\ket{\psi}_{\text{L}}\stackrel{QP}{\rightarrow}(\alpha\ket{R}_{\text{p}}+\beta\ket{L}_{\text{p}})\ket{0}_{{\text{o}}}=\ket{\psi}_{\text{p}}\ket{0}_{{\text{o}}},
\label{decoding}
\end{equation}
which can then be coupled into a single mode fiber and analyzed in
polarization using standard methods. The measurement device is sketched in Fig.\
\ref{fig:2}\textbf{d}. Notice that, again from the linearity of quantum mechanics, the encoding and decoding transformations \eqref{generation} and \eqref{decoding} hold even if the polarization state is part of some larger entangled state. In addition, an outstanding feature of the $q$-plate is that it realizes the polarization-transverse-mode coupling in a single compact device that requires no interferometric stability, therefore providing the scheme with a built-in robustness.

Our first step was to experimentally verify that the encoding/decoding
apparatus works properly in the case of stationary aligned reference frames. We prepared the input photon in one of the polarization states $\ket{H}, \ket{V}, \ket{R}, \ket{L}$, or
$\ket{\pm}=(\ket{H}\pm\ket{V})/\sqrt{2}$. The qubit
was then mapped by a first QP into the rotationally invariant 
encoding, transmitted through free space to the measurement stage, then decoded back to polarization by
a second QP, and finally analyzed in polarization using a set of waveplates and a polarizing beam splitter (PBS). The average measured
fidelity with the input states was $F=(98\pm1)\%$, indicating that the devices work near perfectly.

\begin{figure*}[t]
\begin{center}
\includegraphics[scale=.4]{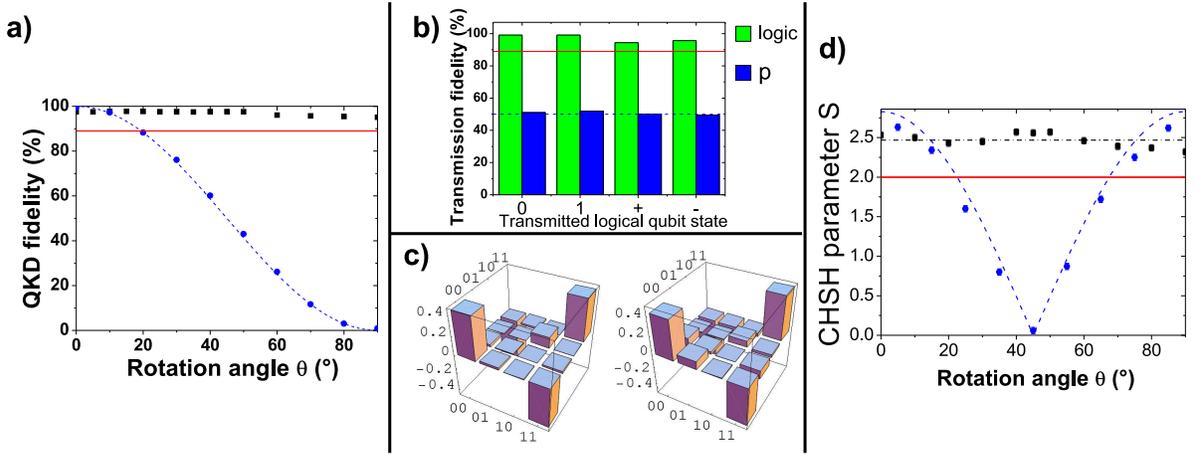}
\end{center}
\caption{{\bf Experimental results of alignment-free quantum
communication tests}. \textbf{a-b)} Measured fidelity
of qubits encoded in the rotationally invariant polarization-OAM space,
in a test of the BB84 quantum key distribution protocol, compared to that of standard polarization-encoded qubits. Panel (a)
shows the fidelity $F_{QKD}$ (square dots) averaged over the four hybrid qubit states used in
the protocol, as a function of the misalignment angle $\theta$ between the transmission and
detection reference frames.  Panel (b) shows the individual fidelity of each of the four states (green bars) observed over the whole QC session including all the different rotation angles probed. The latter accounts for the general situation where the misalignment could vary randomly between transmitted photons. In both panels, the blue dots/bars and dashed lines give respectively the measured and theoretically-calculated fidelity for the standard case of polarization encoding.  The red line delineates the QKD
security threshold. \textbf{c)} Quantum state tomography of the entangled state of hybrid qubits distributed between Alice and Bob,
for the case of aligned reference frames (left) and for a misalignment of $\theta=45^\circ$ (right).  In both cases, only the real part of the density matrices is shown, as the imaginary part is negligible. \textbf{d)} CHSH parameter $S$ (square dots) in experimental non-locality tests on photon pairs entangled in the rotationally invariant qubit space, as a function of the relative misalignment $\theta$ between Alice's and Bob's frames. The red line is the LHV bound. The blue dots represent the measured values of $S$  for the bare polarization-entangled states without the logical protection. Whereas the blue dashed line is the theoretically-calculated maximal CHSH parameter that would be obtained with pure maximally-entangled polarization states. The dot-dashed line in turn represents the overall CHSH value $S$ of the entire test taking into account all the experimental runs with different $\theta$. In all data points, the error bars resulting from Poissonian statistics are smaller than the
symbols.} \label{fig:3}
\end{figure*}
%%%%%%%%%%%%%%%%%%%%%%%%%%%%%%%%%%%%%%%%%%%%%%%%%%%%%%%%
\paragraph{{\bf Alignment-free quantum key distribution.}}
To experimentally demonstrate that the present QC setup works well
for arbitrary relative alignment of Alice and Bob's transverse
reference frames, we mounted the $q$-plate, waveplates, PBS, and optical fiber coupler in a compact and robust detection
stage that can be freely rotated by any angle $\theta$ around the
light propagation axis, as shown in Fig.\ \ref{fig:2}\textbf{d},\textbf{e}. Then, using heralded single photons, and for different angles $\theta$, we encoded, transmitted, and decoded, the four hybrid-qubit states required for the BB84 QKD protocol \cite{Benn84}: $\ket{0}_{\text{L}}, \ket{1}_{\text{L}}$, and $\ket{\pm}_{\text{L}}=(\ket{0}_{\text{L}} \pm \ket{1}_{\text{L}})/\sqrt{2}$. We quantified the potential of our setup for QKD by measuring the fidelities of the states prepared and measured with the ideal ones, as well as the qubit error rates (QBERs) \cite{Scarani09} $\epsilon_{0_{\text{L}}/1_{\text{L}}}$ and $\epsilon_{+_{\text{L}}/-_{\text{L}}}$ for the logical bases $\{\ket{0}_{\text{L}},\ket{1}_{\text{L}}\}$ and $\{\ket{+}_{\text{L}},\ket{-}_{\text{L}}\}$, respectively. The experimental results are reported in Fig.\ \ref{fig:3}\textbf{a,b}. 

\par Fig.\ \ref{fig:3}\textbf{a} shows the average fidelity $F_{QKD}$ over the four states, as a function of $\theta$. $F_{QKD}$ is constantly above the value $F_{\mathrm{T}}=89\%$, which corresponds to the well-known Shor-Preskill security proof threshold \cite{Shor&Preskill}. Above this, under the usual assumptions that  Alice's source emits (logical) qubits, Bob's detectors perform (logical) qubit measurements, and there is no basis-dependent flaw in Alice's and Bob's systems \cite{Gottesman}, unconditional security can be guaranteed. In contrast, the fidelity attained using polarization-encoded qubits falls below the security bound for angles
$\theta>20^{\circ}$, even in the ideal noiseless case (blue dashed line). Fig.\ \ref{fig:3}\textbf{b} in turn shows the fidelity for each state, obtained by uniformly mixing the data over all measured angles $\theta$. Again, all the individual-state fidelities are consistently larger than the security threshold. Indeed, the measured QBERs, estimated as $\text{QBER}=1-F$, were $\epsilon_{0_{\text{L}}/1_{\text{L}}}=(0.65\pm0.09)\%$ and $\epsilon_{+_{\text{L}}/-_{\text{L}}}=(4.1\pm0.2)\%$, from which we expect a high secret-key fraction $r=(70\pm 1)\%$ \cite{Scarani09}. 

%%%%%%%%%%%%%%%%%%%%%%%%%%%%%%%%%%%%%%%%%%%%%%%%%%%%%%%%%%%%%%%
\paragraph{{\bf Alignment-free entanglement distribution}.} 
Fig.\ \ref{fig:2}\textbf{c} shows the experimental setup for the generation of entangled states of
rotationally invariant qubits (see figure caption for symbol definitions): A BBO crystal cut for type-II phase matching
was pumped by the second harmonic of a Ti:Sapphire mode-locked laser
beam, generating photon pairs via
spontaneous parametric fluorescence. The pairs are produced at the
degenerate wavelength $\lambda=795$ nm, with a bandwidth
$\Delta\lambda=3$ nm, as determined by interference filters. The
photon pair is prepared in the polarization entangled state
$\ket{\phi^-}^{AB}_\text{p}=\frac{1}{\sqrt{2}}(\ket{R}_{\text{p}}^A\ket{R}_{\text{p}}^B-\ket{L}_{\text{p}}^A\ket{L}_{\text{p}}^B)$, where the superscripts $A$ and $B$ refer to Alice's and Bob's photons, respectively.
The photons are coupled into single mode fibers (SM) that select only states with zero OAM (TEM$_{00}$ mode). A QP  at the output of each fiber transforms the polarization-entangled state to the rotationally invariant entangled state:
\begin{eqnarray}
\ket{\phi^-}^{AB}_{\text{p}} &\stackrel{QPs}{\rightarrow} \frac{1}{\sqrt{2}}(\ket{0}^A_{\text{L}}\ket{0}^B_{\text{L}}-\ket{1}^A_{\text{L}}\ket{1}^B_{\text{L}})  = \ket{\phi^-}^{AB}_{\text{L}}.
\label{entangled}
\end{eqnarray}

\par To verify the generation of hybrid entanglement, we
performed quantum state tomography of the experimental density matrix
$\rho_{\text{L}}^{AB}$ measured without misalignment ($\theta=0$). The tomographically reconstructed matrix, in the  basis $\{\ket{0}^A_{\text{L}}\ket{0}^B_{\text{L}},\ \ket{0}^A_{\text{L}}\ket{1}^B_{\text{L}},\ \ket{1}^A_{\text{L}}\ket{0}^B_{\text{L}},\ \ket{1}^A_{\text{L}}\ket{1}^B_{\text{L}}\}$, is  shown in Fig.\ \ref{fig:3}\textbf{c} (left side). The fidelity with the experimental polarization entangled state $\rho_{\text{p}}^{AB}$ input to the encoder is $F_{0}(\rho_{\text{L}}^{AB},\rho_{\text{p}}^{AB})=(93\pm 1)\%$, while the entanglement of $\rho_{\text{L}}^{AB}$, as given by the
concurrence, is $C=(0.85\pm 0.03)$. As a first test on the rotational invariance of the state produced, we repeated the  tomographic reconstruction with Alice's measurement stage rotated by $\theta=45^\circ$.  The corresponding reconstruction is shown in Fig.\ \ref{fig:3}\textbf{c} (right side). The fidelity with $\rho_{\text{p}}^{AB}$ is $F_{45}(\rho_{\text{L}}^{AB},\rho_{\text{p}}^{AB})=(96\pm 1)\%$, and the concurrence is $C=(0.84\pm 0.03)$, consistent with the case $\theta=0$.  This indicates that our entanglement distribution scheme is immune to relative misalignments of Alice and Bob.
%%%%%%%%%%%%%%%%%%%%%%%%%%%%%%%%%%%%%%%%%%%%%%%%%%%%%%%%%%%%%%%%%%%%%%%%
\paragraph{{\bf Alignment-free quantum non-locality.}} With the hybrid entangled state \eqref{entangled}, we performed a violation of the CHSH  inequality $S=|E(a_0,b_0)+E(a_1,b_0)+E(a_0,b_1)-E(a_1,b_1)|\leq 2$ in an alignment-free setting. In the inequality, $a_x$ and $b_y$, with possible values 0 or 1, are the outcomes of Alice's and Bob's measurement settings $x$ and $y$, respectively, with $x$ and $y$ equal to 0 or 1. Correlators $E(a_x,b_y)=\langle (-1)^{a_x+b_y} \rangle$, with $\langle  \rangle$ the statistical average, quantify the fraction of events where Alice's and Bob's outcomes are observed to coincide. Any local hidden variable (LHV) model satisfies the inequality \cite{Clau69}. For the rotationally invariant quantum violation of the inequality we chose the following hybrid measurement bases: $\{\ket{0}_{\text{L}},\ket{1}_{\text{L}}\}$ and $\{\ket{+}_{\text{L}},\ket{-}_{\text{L}}\}$, corresponding to Alice's settings $x=0$ and $x=1$, respectively, and $\{\cos\frac{\pi}{8}\ket{0}_{\text{L}}+\sin\frac{\pi}{8}\ket{1}_{\text{L}},\; -\sin\frac{\pi}{8}\ket{0}_{\text{L}}+\cos\frac{\pi}{8}\ket{1}_{\text{L}}\}$ and $\{\sin\frac{\pi}{8}\ket{0}_{\text{L}}+\cos\frac{\pi}{8}\ket{1}_{\text{L}},\; -\cos\frac{\pi}{8}\ket{0}_{\text{L}}+\sin\frac{\pi}{8}\ket{1}_{\text{L}}\}$, corresponding to Bob's settings $y=0$ and $y=1$, respectively. Fig.\ \ref{fig:3}\textbf{d} reports the measured CHSH parameter $S$ versus the rotation angle $\theta$ 
of Alice's measurement frame. The figure shows that the LHV bound is violated for all angles, in striking contrast with the experimental polarization state $\rho_{\text{p}}^{AB}$ (blue circles), or even with the ideal maximally-entangled polarization state $\ket{\phi^-}^{AB}_\text{p}$ (blue dashed line). For the logically-encoded states, we mixed the data of all different values of $\theta$ to test the violation's immunity to arbitrarily-varying frame orientation, obtaining a value of $S=(2.47\pm0.01)>2$. This alignment-free extraction of non-local correlations reconfirms the rotational invariance of the quantum resources created here.
%%%%%%%%%%%%%%%%%%%%%%%%%%%%%%%%%%%%%%%%%%%%%%%%%%%%%%%%%%%%%%%%%%%%%%%%%%%%%%%%%%%%%

\paragraph{\bf Robustness of hybrid rotational-invariant qubits.} A remarkable feature of our polarization-OAM hybrid-encoding QC scheme is that it turns out to be very robust against the spatial-mode perturbations arising in beam misalignments around axes other than the optical one and atmospheric turbulence effects. Such robustness appears at first glance counterintuitive, since the encoding  involves the use of orbital angular momentum, which is quite sensitive to all the above-mentioned spatial perturbations \cite{Pate05,Giov11}. The main reason of such robustness is that the OAM spread induced by spatial-mode perturbations is neutralized by the polarization degree of freedom, which is in contrast very robust against those spatial-mode perturbations. This allows to filter out, in the receiving unit, most components of the state that would otherwise decrease the fidelities. That is, the particular decoding setup used intrinsically implements an 
effective quantum error-correction procedure that (detects and) discards (but does not correct) all states outside the logical subspace.

\par Indeed, spatial-mode perturbations will alter a generic hybrid qubit $\alpha\ket{R}_p\ket{l}_o+\beta\ket{L}_p\ket{r}_o$, transforming it into the following photon state:
\begin{equation}
\label{newequation}
\sum_{m}[c(+1,m)\alpha\ket{R}_p\ket{m}_o+c(-1,m)\beta\ket{L}_p\ket{m}_o],
\end{equation}
where $\ket{m}_o$ denotes a generic mode with OAM eigenvalue $m\hbar$ and $c(k,m)$ are the probability amplitudes for the photon OAM to be shifted from $k\hbar$ to $m\hbar$, due to the perturbation. The presence of the unperturbed polarization leads the state \eqref{newequation} to be projected, in the final measurement stage, onto its components with $k= m$, thus filtering exclusively the desired logical subspace. In addition, the residual ``diagonal'' effect can be shown to disappear, i.e. $c(+1,+1)=c(-1,-1)$, for a large class of perturbations, including rotations about axes orthogonal to the propagation direction. We have experimentally verified the robustness of hybrid qubits for different perturbations, observing state fidelities that remain high (in particular, above the unconditional security bound). Further details and calculations are reported in the Supplementary Information.
%%%%%%%%%%%%%%%%%%%%%%%%%%%%%%%%%%%%%%%%%%%%%%%%%%%%%%%%%%%%%%%%%%%%%%%%
\par{\bf Conclusion and discussion}. Quantum communication plays a fundamental role in the modern view of quantum physics and opens the possibility of a variety of technological applications. Uncontrolled reference-frame misalignments limit QC, as they turn the transmitted quantum messages into noisy, classical ones. Here, we reported the development of a robust and compact toolbox for the efficient encoding and decoding of quantum information into  single-photon states that are invariant under arbitrary rotations around the optical axis.  With this, all concerns on relative axis-orientation during quantum transmissions reduce simply to the basic requirement of establishing an optical link. 

\par Rotational invariance is achieved by exploiting decoherence-free subspaces spanned by hybrid polarization-orbital angular momentum entangled states.   We experimentally showed the efficacy of these states through the feasibility-demonstration of
a cryptographic-key distribution protocol, distribution of entanglement, and violation of a Bell
inequality, all in alignment-free settings. Importantly,  as for what cryptographic security concerns, our scheme does not introduce loopholes other than those already present in any photonic experiment with conventional encodings. We also emphasize that, even though the states used are themselves invariant only under rotations about the propagation axis, the scheme resists misalignments around other directions too. This is due to a filtering mechanism intrinsic to our universal-decoder setup, which maps errors originating from beam rotations around axes other than the optical link, as well as other spatial perturbations, into signal loses instead of infidelity (see Supplementary Information).

\par As mentioned, recent interesting alignment-free approaches for QKD \cite{Laing10} and to extract non-local correlations  \cite{Lian10, Shad11,Wall11} have been put forward. These however require that the relative axis-orientations, though unknown, stay approximately static throughout the quantum data exchange session (see the Supplementary Information for details). Remarkably, in contrast, our rotational-invariance protection works even if the relative orientations vary arbitrarily from measurement to measurement. Moreover, another important feature of the present scheme is that it does not restrict to non-locality and QKD but enables also fully general QC protocols, all misalignment-immune \cite{Aoli07}. These include for instance quantum teleportation, dense coding, and entanglement swapping, the main ingredient of quantum repeaters. 

Finally, our scheme should find applications in the forthcoming experiments \cite{Rari02,Aspe03,Vill08,Bona09} on long-distance satellite-based quantum communication. There, apart from misalignments, other issues may impose serious obstacles too, as precise satellite laser-tracking, collection efficiencies, or finite-size effects (for QKD). However, immunity against arbitrarily-varying transverse relative orientations not only solves for misalignments but also relaxes requirements on the repetition rates needed to overcome finite-size effects (see Supplementary Information).

%%%%%%%%%%%%%%%%%%%%%%%%%%%%%%%%%%%%%%%%%%%%%%%%%%%%%%%%%%%%%%%%%%%%%%%%
\par{\bf Acknowledgements}. This work was supported by the FET-Open Program, within the 7$^{th}$
Framework Programme of the European Commission under Grant No.
255914, PHORBITECH, FIRB-Futuro in Ricerca (HYTEQ), the Brazilian funding agencies FAPERJ, CNPq and the INCT-Informa\c{c}\~ao Qu\^antica, and the Spanish Juan de la Cierva foundation.

\end{document}

\title{Supplementary Information: Complete experimental toolbox for alignment-free quantum communication}
\author{Vincenzo D'Ambrosio}
\author{Eleonora Nagali}
\affiliation{Dipartimento di Fisica, Sapienza Universit\`{a} di Roma, Roma 00185, Italy}
\author{Stephen P. Walborn}
\affiliation{Instituto de F\'isica, Universidade Federal do Rio de Janeiro, Rio de Janeiro, RJ 21941-972, Brazil}
\author{Leandro Aolita}
\affiliation{ICFO-Institut de Ci\`{e}ncies Fot\`{o}niques, Av. Carl Friedrich Gauss 3, 08860 Castelldefels (Barcelona), Spain}
\author{Sergei Slussarenko}
\affiliation{Dipartimento di Scienze Fisiche, Universit\`{a} di Napoli ``Federico II'',
Compl.\ Univ.\ di Monte S. Angelo, 80126 Napoli, Italy}
\author{Lorenzo Marrucci}
\affiliation{Dipartimento di Scienze Fisiche, Universit\`{a} di Napoli ``Federico II'',
Compl.\ Univ.\ di Monte S. Angelo, 80126 Napoli, Italy}
\affiliation{CNR-SPIN, Complesso Universitario di Monte S. Angelo, 80126 Napoli, Italy}
\author{Fabio Sciarrino}
\email{fabio.sciarrino@uniroma1.it}
\affiliation{Dipartimento di Fisica, Sapienza Universit\`{a} di Roma, Roma 00185, Italy}
\maketitle

In the first part of this Supplementary Information we show, both theoretically (Section I) and experimentally (Section II), that the rotational-invariant spin-orbit hybrid encoding of the photonic qubits that we propose for alignment-free quantum communication (QC) is highly resistant to a wide class of perturbations affecting the spatial mode. The latter for example include imperfect beam alignment (e.g., displacement, tilt), beam partial obstruction or masking due to finite optical apertures of the optical setup, and atmospheric turbulence. By resistant we mean here that the qubit fidelity remains very high (and hence the error rates low), even if the transmission efficiency is depressed or the photon losses increased.

This high resistance is counterintuitive, since the hybrid encoding of the photonic qubits involves the use of orbital angular momentum (OAM), which is known to be quite sensitive to all the above-mentioned spatial perturbations (although significant progresses in pure OAM-based classical and quantum communication through the atmosphere have been recently reported \cite{Djord11,Zhao12,Pors11}). As we shall see, the underlying physical reason is that our decoding procedure from the hybrid logical space back to polarization features an intrinsic error-correction mechanism that filters out most components of the state that would introduce qubit alterations. This in turn relies on the use of the polarization degree of freedom, which is essentially unaffected by the spatial-mode perturbations.

Then, in Section III of this Supplementary Information, we briefly review other previously proposed approaches to quantum communication without a shared reference frame (SRF), comparing their main properties with those of our approach.
%%%%%%%%%%%%%%%%%%%%%%%%%%%%%%%%%%%%%%%%%%%%%%%%%%%%%%%%%%%%%%%%%%%%%%

\section{Resistance of hybrid qubits to spatial perturbations: theory}

To fully describe the action of perturbations of the spatial mode, we need to consider both the azimuthal quantum number $m$, corresponding to the OAM eigenvalue in units of $\hbar$, and a radial quantum number $p$. These two numbers can be for example defined as in the case of Laguerre-Gauss beams (LG). However, the radial basis utilized is not constrained to be that of LG beams, and our treatment is general in this respect. In this Section, we denote with $\ket{m,p}$ the spatial mode and with $\ket{P,m,p}$ the entire photon quantum state, including the polarization state $P$ (e.g., $L/R$ for left/right circular polarization, or $H/V$ for horizontal/vertical linear polarization).

Let us consider the action of generic spatial-mode perturbations, as defined by the following transformation laws:
\begin{equation}
\ket{m,p} \rightarrow \sum_{p',m'}C_{m,m';p,p'}\ket{m',p'} \label{perturb}
\end{equation}
where $C_{m,m';p,p'}$ are suitable complex-valued coefficients [ignoring the radial modes, they correspond to the $c(m,m')$ coefficients used in Eq. (6) of the main Article]. Notice that we assume here that the perturbation does not affect the polarization state.

Next, we need to specify the full behavior of the q-plate (with $q=1/2$) including the radial mode, which can be described by the following transformation laws:
\begin{eqnarray}
\ket{L,m,p} &\rightarrow& \sum_{p'}Q_{|m|,|m+1|;p,p'}\ket{R,m+1,p'} \nonumber\\
\ket{R,m,p} &\rightarrow& \sum_{p'}Q_{|m|,|m-1|;p,p'}\ket{L,m-1,p'} 
\end{eqnarray}
where $Q_{m,m';p,p'}$ are coefficients which do not depend on the sign of $m$ and $m'$, owing to the mirror symmetry of the q-plate pattern. These coefficients can be also given explicit analytical expressions in a given radial basis (e.g., the LG one), but these expressions are not needed for our purposes here.

Let us now consider a generic input polarization-encoded qubit photon in Gaussian mode TEM$_{00}$ ($m=0, p=0$):
\begin{equation}
\ket{\psi}_P=\alpha\ket{R,0,0}+\beta\ket{L,0,0}.
\end{equation}
After the q-plate, this photon is converted into the following rotation-invariant hybrid state (corresponding to the logical qubit):
\begin{equation}
\ket{\psi}_L=\sum_{p}Q_{0,1;0,p}\left(\alpha\ket{L,-1,p}+\beta\ket{R,+1,p}\right).
\end{equation}
Notice that, for $p=0$, the last state reduces to the right-hand side of transformation (3) in the main text. We now consider the effect of the perturbation (\ref{perturb}) on the photon, which is thus transformed into the following state:
\begin{eqnarray}
\ket{\psi'}_L&=&\sum_{p,m,p'}Q_{0,1;0,p}\left(\alpha C_{-1,m;p,p'}\ket{L,m,p'} \right.\nonumber\\
&& \left. +\beta C_{+1,m;p,p'}\ket{R,m,p'}\right).
\end{eqnarray}
Next, we apply to this perturbed state the transformations used in the decoding unit needed to return back to the polarization encoding (q-plate transformation):
\begin{eqnarray}
\ket{\psi'}_P&=&\sum_{p,m,p',p''}Q_{0,1;0,p} \nonumber\\
&\times& \left(\alpha Q_{|m|,|m+1|;p',p''}C_{-1,m;p,p'}\ket{R,m+1,p''}\right. \\
&&\left. +\beta Q_{|m|,|m-1|;p',p''}C_{+1,m;p,p'}\ket{L,m-1,p''}\right). \nonumber
\end{eqnarray}
Before detection, the photon is coupled to a single-mode fiber, which performs a spatial filtering that allows only the $m=0, p=0$ states to pass through to the detector, thus eliminating all other modes:
\begin{eqnarray}
\ket{\psi''}_P&=&\sum_{p,p'}Q_{0,1;0,p}Q_{1,0;p',0} \\
&\times&\left(\alpha C_{-1,-1;p,p'}\ket{R,0,0}+\beta C_{+1,+1;p,p'}\ket{L,0,0}\right). \nonumber
\end{eqnarray}
The latter equation shows that the qubit state will be unaffected, except for a global amplitude and phase factor, if and only if the following equality holds true:
\begin{equation}
C_{-1,-1;p,p'} = C_{+1,+1;p,p'}, \label{pertcond}
\end{equation}
for all values of the radial indices $p,p'$. In other words, any spatial-mode perturbation which satisfies Eq. (\ref{pertcond}) will not alter the qubit transmission fidelity, although it may affect the transmission efficiency by increasing the photon losses.

In particular, every beam transformation that is mirror-symmetric with respect to a plane containing the initial beam axis will be symmetrical in the sign of OAM and hence will automatically satisfy Eq. (\ref{pertcond}). For example a beam parallel displacement, tilt, elliptical deformation, or aperturing with a circular iris (even if off center) or a half-plane mask (knife-edge) all have this symmetry and hence will not affect the transmission fidelity. An axial misalignment between the transmitting and receiving communication units is equivalent to a beam translation and/or tilt and can be treated analogously. Only symmetry-breaking combinations of two or more of the above effects may affect the fidelity. For example a beam displacement combined with a beam tilt, if they are not along the same plane, will break the mirror symmetry and hence might introduce some degree of qubit alteration. Also, the main optical effects arising from atmospheric turbulence, such as beam wandering and spreading are mirror-symmetric, so that the extent of qubit alteration is anticipated to be much reduced in our communication scheme than in the case of pure OAM communication.

Another important class of transformations which satisfies Eq. (\ref{pertcond}) is that defined mathematically by pure multiplicative factors acting on the optical field, e.g. the transformations arising from crossing any arbitrary inhomogeneous medium that is thin as compared with the Rayleigh length. It is easy to verify that these will be described by coefficients $C_{m,m';p,p'}$ which depend on the difference $m-m'$ and on the absolute values $|m|$ and $|m'|$ (besides $p$ and $p'$), so that Eq. (\ref{pertcond}) is automatically satisfied. Weak turbulence, introducing only pure phase wavefront distortions, falls within this class of transformations and is therefore predicted to leave the qubit fidelity unmodified \cite{Pate05}. If we now consider the fact that light propagation in homogeneous media leaves the various OAM components constant, we conclude that  Eq. (\ref{pertcond}) is satisfied even if the turbulent medium is followed and/or preceded by a long-distance free-space propagation, as in the case of earth to satellite (and vice versa) communication through the atmosphere.

\subsection{Parallel beam displacement}
The effect of beam translation for a LG beam with initial $p=0$ and $m=1$ was treated in \cite{Vasne05} [equations (4)-(7)]. We can generalize the reported result to the case of initial $p=0$ and $m=\pm1$, obtaining the following expression for the translated beam in polar coordinates $\rho,\varphi$:
\begin{eqnarray}
E(\rho,\varphi)&=&\frac{A}{w_0}\left(\rho e^{\pm i\varphi}-\delta e^{\pm i\theta}\right) e^{-\frac{\rho^2+\delta^2}{w_0^2}}\nonumber\\
&&\times\sum_{m=-\infty}^{+\infty}I_m\left(\frac{2\rho\delta}{w_0^2} \right) e^{im(\varphi-\theta)}
\end{eqnarray}
where $\delta$ and $\theta$ are the polar coordinates of the displacement vector in the plane orthogonal to the beam axis $z$, $w_0$ is the beam waist, $A$ a normalization constant, and $I_m$ are the modified Bessel functions of the first kind.

By projection of the latter expression on a LG mode with $p'=0$ and $m'=m=\pm1$, we obtain the following transformation coefficients:
\begin{eqnarray}
C_{m,m';0,0}&=&\frac{2\pi A^2}{w_0^2}\int_0^{\infty}\left\{ \rho^3 e^{-\frac{2\rho^2+\delta^2}{w_0^2}} \right. \nonumber\\
&& \left. \left[ \rho I_0\left(\frac{2\rho\delta}{w_0^2}\right) -\delta I_1\left(\frac{2\rho\delta}{w_0^2}\right) \right]  \right\} d\rho
\end{eqnarray}
which are independent of the sign of $m=m'$ and hence satisfy Eq. (\ref{pertcond}). A similar, though more complex, analysis can be carried out for arbitrary $p$ and $p'$.

\subsection{Beam tilting}
We generalize the results given in \cite{Vasne05} [equations (11)-(13)], obtaining the following expression for the transformation coefficients representing the effect of beam tilt on LG beam having $p=p'=0$ and $m=m'=\pm1$:
\begin{equation}
C_{m,m';0,0}=\frac{2\pi A^2}{w_0^2}\int_0^{\infty}\rho^3 e^{-\frac{2\rho^2}{w_0^2}} J_0(\alpha\rho)  d\rho
\end{equation}
where $\alpha=k\sin\gamma$, with $k$ the beam wavenumber and $\gamma$ the tilt angle. The tilt azimuthal angle $\eta$ is irrelevant here, and Eq. (\ref{pertcond}) is satisfied.

\subsection{Combination of beam tilt and displacement}
We now generalize the results given in \cite{Vasne05} [equations (14)-(17)], obtaining the following expression for the transformation coefficients representing the combined effect of beam tilt and displacement on LG beam having $p=p'=0$ and $m=m'=\pm1$:
\begin{equation}
C_{m,m';0,0} = \frac{2\pi A^2}{w_0^2}\int_0^{\infty}\rho^3 e^{-\frac{2\rho^2+\delta^2}{w_0^2}} \left[\rho S_0(\rho) - \delta S_1(\rho) \right]  d\rho
\end{equation}
where
\begin{equation}
S_0(\rho) = \sum_{n=-\infty}^{+\infty} I_{|n|} \left( \frac{2\rho\delta}{w_0^2}\right) J_n(\alpha\rho) e^{in(\theta-\eta+\pi/2)}
\end{equation}
and
\begin{equation}
S_1(\rho) = \sum_{n=-\infty}^{+\infty} I_{|n-m|} \left( \frac{2\rho\delta}{w_0^2}\right) J_n(\alpha\rho) e^{in(\theta-\eta+\pi/2)}
\end{equation}
On inspection, we find that this result satisfies Eq. (\ref{pertcond}) if $\theta=\eta$ (or $\theta=\eta\pm\pi$), i.e.\ tilt and displacement occur in the same (or opposite) azimuthal direction. This is consistent with the general analysis based on the mirror symmetry of the transformation, which is broken if $\theta\neq\eta$ and $\theta\neq\eta\pm\pi$.
%%%%%%%%%%%%%%%%%%%%%%%%%%%%%%%%%%%%%%%%%%%%%%%%%%%%%%%%%%%%%%%%%%%%%%

\section{Resistance of hybrid qubits to spatial perturbations: experiments}
We experimentally test the resistance of the proposed alignment-free QC scheme to a set of spatial-mode perturbations, in order to verify that the theory presented in the previous section is valid also in the presence of the unavoidable imperfections of any experimental setup. In all cases, we also compare the transmission fidelity of rotational-invariant hybrid qubits with that of pure OAM-encoded qubits.
\begin{figure}
\begin{center}
\includegraphics[scale=0.3]{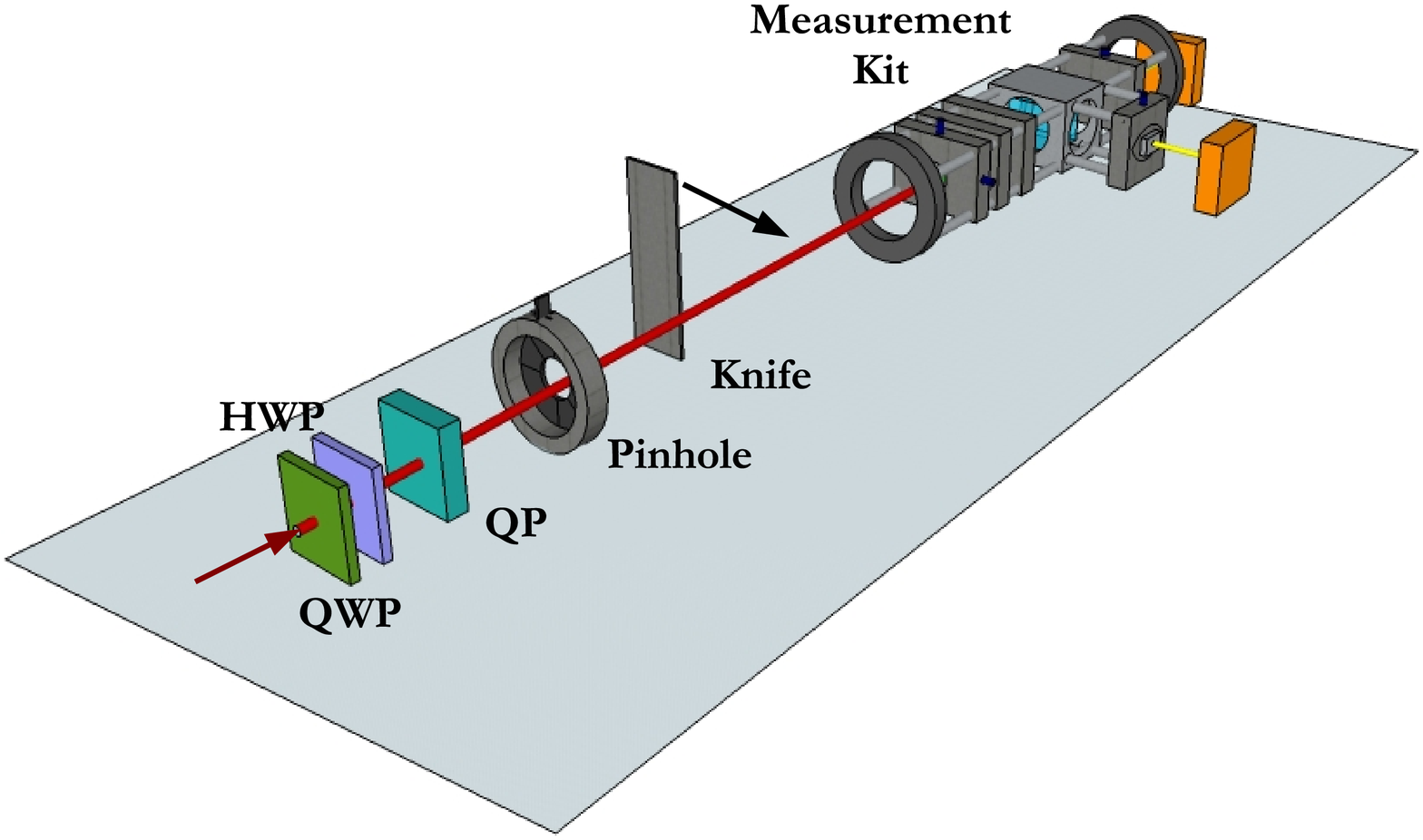}
\caption{Experimental setup adopted for the tests on the resistance of the rotational invariant hybrid photonic qubits to spatial-mode perturbations. In the schematics, we reported both the circular aperture (pinhole) and the half-plane obstruction (a movable knife-edge) that can alter the transmission of the qubits.}
 \label{fig5si}
\end{center}
\end{figure}
\begin{figure*}
\begin{center}
\includegraphics[scale=0.45]{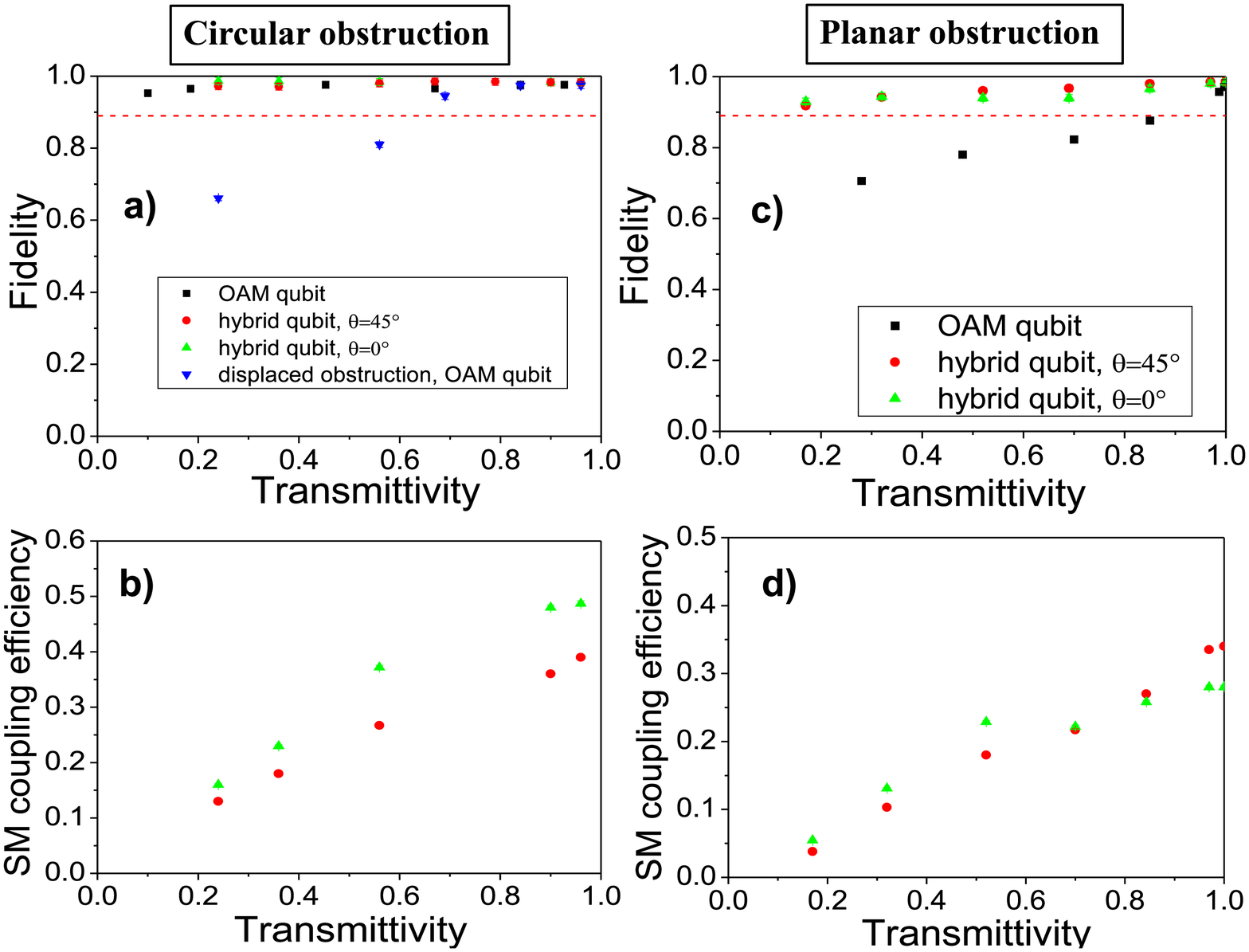}
\caption{Experimental resistance of rotational invariant hybrid qubits to beam perturbations, compared with the case of pure OAM qubits. \textbf{a)} Circular aperture. Average fidelity of pure OAM qubits (black squares), hybrid qubits for a measurement stage rotated at an angle $\theta=0^\circ$ (green triangle) and at an angle $\theta=45^{\circ}$ (red circles) with respect to the transmitting unit. The blue triangles refer to the pure OAM qubits case, when the circular aperture is displaced off the beam axis by $5\%$ of the beam waist (the hybrid qubit behavior in the latter case was essentially indistinguishable from the centered aperture case). \textbf{b)} Transmission efficiency determined by single-mode (SM) fiber-coupling efficiency after a circular aperture of varying radius, in the case of hybrid qubits only. \textbf{c)} Half-plane aperture. Average fidelities for hybrid qubits at $\theta=0^{\circ}$ (green triangle) and $\theta=45^\circ$ (red circles). Black squares are the corresponding results for pure OAM qubits. \textbf{d)} Single mode (SM) coupling efficiency after a movable half-plane aperture.}
 \label{fig6si}
\end{center}
\end{figure*}

\subsection{Experimental setup}
The input photon pairs are generated via spontaneous parametric fluorescence in a $\beta$-barium borate crystal cut for type-II phase matching, pumped by the second harmonic of a Ti:Sa mode-locked laser beam with repetition rate equal to $76$MHz. The generated photons have wavelength $\lambda=795$
nm, and spectral bandwidth $\Delta\lambda=3$ nm, as determined by two interference filters (IF). The spatial and temporal walk-off is compensated by inserting a $\frac{\lambda }{2}$ waveplate and a $0.75$ mm thick BBO crystal ($C$) on each output mode \cite{Kwia95}. The detected coincidence rate of the source is $C_{source}=8$ kHz. The photons are delivered to the setup via single-mode fibers, in order to define their transverse spatial mode to a pure TEM$_{00}$, corresponding to OAM $m=0$.  After the fiber output, two waveplates compensate the polarization rotation introduced by the fiber. As described in the main text, the maximally entangled state $\ket{\phi^-}^{AB}_P$ generated by the source encoded in the polarization degree of freedom is transformed to the rotationally invariant entangled state $\ket{\phi^-}^{AB}_L$ by inserting on each arm of the setup a q-plate with topological charge $q=1/2$. The conversion efficiency of all the q-plates employed in the experiment has been optimized by controlling the electric field applied to the device \cite{Damb12}, so that the q-plates generate the rotational-invariant hybrid qubits with a mean efficiency equal to $(94\pm 2)\%$. 

We note in passing that in principle the q-plate used in our work for converting the polarization encoding of qubits into the hybrid rotation-invariant one could be also replaced with a complex arrangement of standard polarization and OAM generation/measurement devices. However they would be significantly less efficient (for example, spatial light modulators typically cannot exceed 40-45\% of efficiency when used for measuring an OAM-encoded qubit in a given basis) and more complex to align. In addition these standard methods would only allow for generating and measuring the hybrid qubit locally, while the q-plates may be used to transmit (without SRF) unkown qubits coming from external remote sources and encoded in polarization. Or, at the receiver site, the q-plate allows using the received qubit in further quantum processing based on polarization encoding, without actually measuring it.

After the encoding process, the photons are sent to the measurement stages described in the main article text: see Fig. 2-\textbf{d,e}. For the analysis of experimental data, we referred to the coincidence counts between detectors $[D_{A1},D_{B1}]$ and $[D_{A2},D_{B2}]$, collected by a coincidence circuit with a gate of $3$ ns. The receiving units are located at a distance of $(60.0\pm0.5)$cm from the transmitting unit, and the mean value of the coincidence events between Alice and Bob detectors reads $30$Hz for the CHSH experiment and $300$Hz for the BB84 measurements. 
For the BB84 experiment, one of the two photons emitted by the source has been coupled to a single mode fiber connected to a single-photon counter modules, thus acting as a trigger on the single photon state generation.

\subsection{Resistance of hybrid qubits to beam perturbations}
We considered transmission through two types of transverse obstructions: a half-plane movable obstruction covering a variable fraction of the transverse mode, and an iris (or pinhole) with variable radius. We have measured the state transmission fidelity $F$ for different input states, at both aligned and $45^{\circ}$-rotated measurement stage with respect to the transmitting unit, and for an increasing disturbance due to the obstruction. The experimental setup used for this test is illustrated in Fig.\ref{fig5si}. We encode different polarization qubits using two waveplates (HWP and QWP), and we convert them to the hybrid encoding using a q-plate. For the purpose of comparison, we can also switch to a pure-OAM encoding by inserting a fixed linear polarizer after the q-plate, so as to erase the polarization content of the qubit. Then the photon was sent through the obstruction and to the receiving unit. By changing the transmittance of the input qubit through the obstruction, varying the aperture of the pinhole or the transverse position of the knife, we measured the  fidelity of the transmitted quantum state.  Thus the lowest transmittivity corresponds to a tiny aperture of the pinhole ($0.2$ times the beam size), or to the almost complete coverage of the beam. All reported experimental fidelities are obtained as averages over the six eigenstates of three mutually unbiased bases, thus providing a good representative of the average fidelity over any input qubit state. The experimental results are reported in Fig.\ref{fig6si}.

We notice that the fidelity of hybrid qubits is independent both of the transmittivity of the circular aperture and of the rotation angle of the measurement kit. Moreover, the fidelity is not affected by the displacement of the pinhole off the beam axis. The average fidelity for all data points in the case of hybrid qubits is $F=(98\pm 1)\%$. For comparison, we tested the resistance of qubits encoded only in the two-dimensional OAM subspace $o_1=\{\ket{+1},\ket{-1}\}$, i.e., the same subspace used for the hybrid encoding. We note that for pure OAM qubits the fidelity is strongly affected by the exact centering of the circular aperture. Indeed when the circular aperture is properly centered, we obtain again a constant trend of state fidelity with mean value $F=(97\pm 1)\%$, as expected because a centered circular aperture does not break the beam cylindrical symmetry and hence does not affect the OAM state \cite{Giov11}. On the other hand, when the center of the pinhole is displaced, the OAM qubit fidelity decreases rapidly as the transmittivity is lowered due to the symmetry breaking, as shown by the blue triangles. 
\begin{figure}
\begin{center}
\includegraphics[scale=0.3]{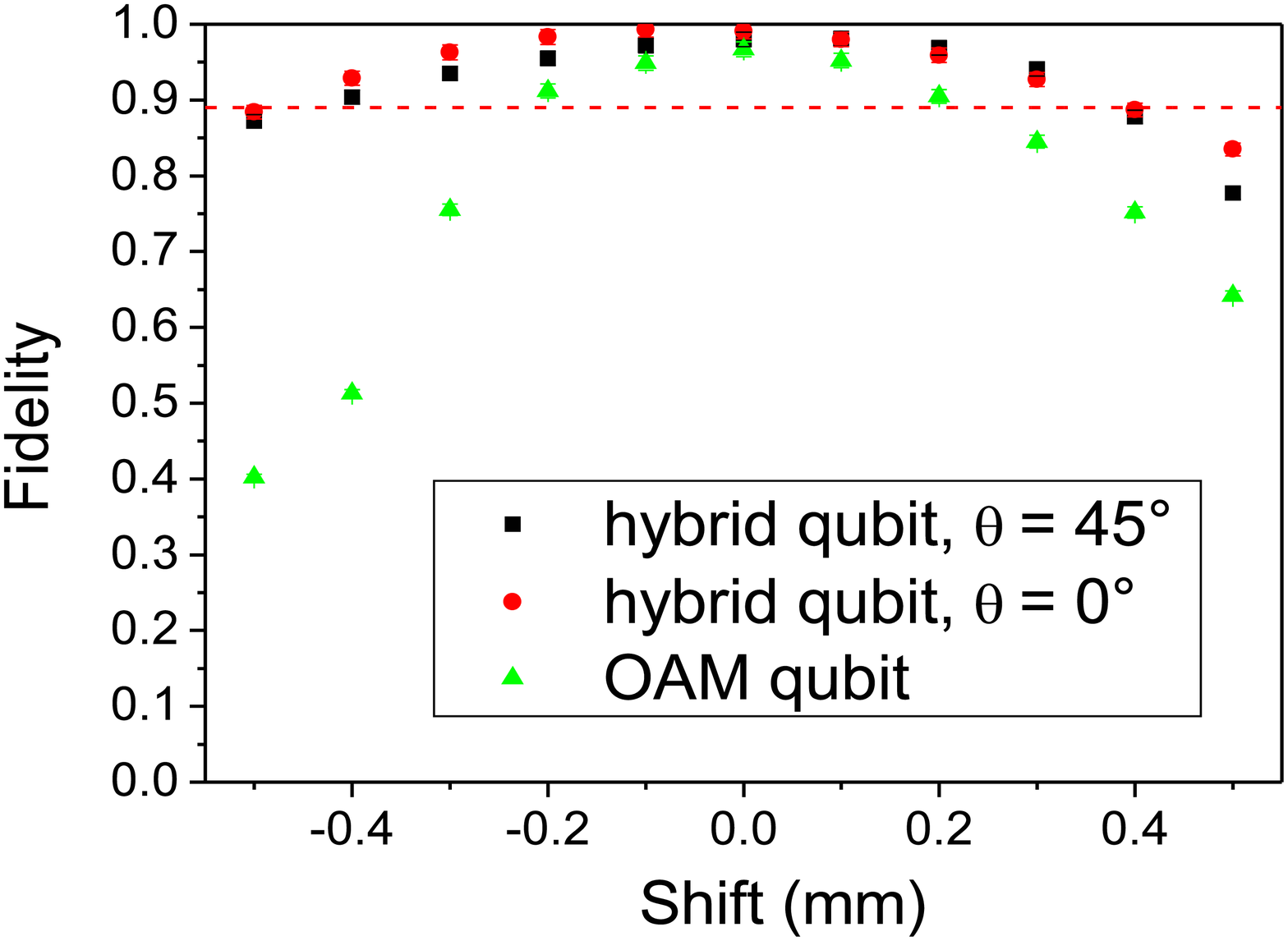}
\caption{Experimental resistance of hybrid qubits to a beam displacement, compared to the case of pure OAM encoding. The beam waist in our experiment is $w_0=(1.0\pm 0.1)$mm.}
 \label{fig7si}
\end{center}
\end{figure}
A similar study has been carried out for the case of a half-plane aperture: Fig.\ref{fig6si}-\textbf{c-d}.  We first estimated the state fidelity for hybrid qubits  with the measurement kit rotated by $\theta=0^\circ$ and $\theta=45^\circ$. Even in this case the resistance of hybrid qubits is confirmed even in a high loss-regime, with a mean value of $F=(96\pm 1)\%$ over all data points. On the other hand, for pure OAM qubits we observe a reduction of the state fidelity as the transmittivity decreases. As shown in \cite{Giov11}, the half-plane obstruction leads to a spread in the OAM spectrum, due to the breaking of the rotational symmetry of the mode. Such effect implies a cross-talk between different OAM contributions, and thus alters the fidelity of pure OAM qubits. Considering hybrid qubits, the presence of polarization implies that the spreading will only involve states that fall outside the hybrid spin-orbit subspace used for the encoding. Such contributions are filtered out in the measurement process, so that the overall number of detected photons decrease, but the state fidelity remains unaffected.

\subsection{Resistance of hybrid qubits to beam displacement and misalignment}
We performed two further tests to verify the hybrid qubit resistance to beam misalignments. First, we performed a communication test while changing the angle of the measurement kit without reoptimizing the alignment of the single mode fiber. This corresponds to introducing small uncontrolled tilt and displacements in the beam during the measurement. Next, we tested the communication fidelity dependence on a controlled beam displacement, for two fixed angles of the measurement stage.

In the first test, we found that the system preserves a good quantum communication fidelity (i.e. above the security threshold) for rotations up to $30^{\circ}$. Above this angle it was necessary to slightly readjust the single mode fiber in order to restore a high fidelity.
Fig.\ \ref{fig7si} shows the behavior of the average communication fidelity when a controlled displacement of the beam with respect to the measurement stage axis is introduced. The hybrid qubit fidelity decreases with displacement, but much slower than for pure OAM encoding.

We believe that this small fidelity reduction effect observed in the last experiment is due to imperfections in the q-plate device. In particular, the central defect in the q-plate, which ideally should be pointlike, has an extension of about 100 $\mu$m in our devices. This introduces a small component of light that is not properly decoded in the measurement stage and therefore introduces a perturbation in the qubit. This effect is usually negligible in the case of well-aligned beams because the defect coincides with the beam vortex, so there is almost no light being affected. On the other hand, when the beam is displaced the defect is overlapped with a bright region of the beam and starts to have a more significant effect. At the same time, the coupling to the fiber of the light crossing the ``good'' regions of the q-plate is reduced, so that the relative role of the defect is enhanced. We are confident that this problem can be solved by using light beams with a larger waist and by improving the manufacturing process of the q-plate.
%%%%%%%%%%%%%%%%%%%%%%%%%%%%%%%%%%%%%%%%%%%%%%%%%%%%%%%%%%%%%%%%%%%%%%

\section{Discussion on previous contributions towards alignment-free quantum communication}
 In this section we briefly review previous experiments and theoretical proposals on misalignment-free quantum communication. 

We begin by prepare-and-measure QKD protocols: In Ref.\ \cite{Laing10}, an ingenious technique was proposed based on circularly polarized single-photon states. This scheme (from now on called the LSRO10 scheme) is suitable for unknown, and possibly misaligned,  relative transverse-axis orientations. However, to bound the knowledge of potential eavesdroppers, a tomographically-complete set of correlations between sender preparations and receiver measurements must be determined \cite{Le11}. 
If the relative misalignment varies in an uncontrolled fashion during the signal-acquisition process, the necessary correlations are smeared out and therefore no security can be guaranteed. For this reason, the scheme is applicable if the misalignment angle $\theta$ varies at slow rate over a time long enough to collect enough signals to overcome finite-sized key effects. 

This has been quantitatively studied in \cite{Sheridan10}, accounting for realistic conditions as non-perfect classical post-processing, QBERs of 5$\%$, and for $\theta$ varying both at constant speed or in a random walk. The authors found for instance that, for secret-key fractions of $r\approx 5\%$, the LSRO10 scheme requires about $10^7$ signals, and this holds only if $\theta$ varies at most $10^{-10}$ (for constant rotation) and $10^{-5}$ (for random-walk rotation) degrees from signal to signal  uninterruptedly throughout the entire signal collection. Our scheme instead allows a BB84 implementation which does not suffer from such restriction, as it is immune to arbitrary variations of $\theta$. 

Considering non-locality tests, recently interesting alignment-free approaches to extract non-local correlations have been put forward ( Refs. \cite{Shad11,Wall11}). These are based on the fact that, even for randomly chosen settings, there is always a finite probability of observing non-locality \cite{Lian10}. However these require that $\theta$ stays fixed throughout the data exchange session. 

Misalignment-immune QC, for the restricted case of a single logical qubit has been previously demonstrated in Ref.\ \cite{Chen06}. This experiment used four physical qubits, realized with the polarization and time-bin degrees of an entangled-photon pair, to encode a logical one. That is, it required a parametric down conversion setup, plus a complex interferometer, to encode a single logical qubit. The main disadvantage of this approach is that, since two photons are used in the encoding, the sensitivity to photon losses increases quadratically.  For example, in a scenario of satellite-to-earth QC, losses may typically be greater than $10^{-9}$ per photon (see Ref.\ \cite{Vill08}).  Thus, two-photon encodings must overcome losses of around $10^{-18}$.  
In addition to losses, the state preparation in the approach of Ref.\ \cite{Chen06} is probabilistic, so that only about 1/3 of the pairs produced are actually used. Moreover, an interferometric setup is sensible to optical-path fluctuations and thus implying non-trivial compensations in an hypothetical distant moving station.
%%%%%%%%%%%%%%%%%%%%%%%%%%%%%%%%%%%%%%%%%%%%%%%%%%%%%%%%%%%%%%%%%%%%%%
\section{Summary}
In the first part of this Supplementary Information we have shown, both theoretically and experimentally, that the presented alignment-free quantum communication scheme is highly resistant to a large class of spatial-mode perturbations. The transmission fidelity of rotational-invariant hybrid qubits has been shown to remain high in the presence of these perturbations, and in particular to be always much higher than in the case of pure OAM-encoded qubits. The transmission fidelity was also proved to be relatively insensitive to beam misalignment, wandering effects, obstructions, etc.  The key feature behind such fidelity robustness is an error-correction mechanism intrinsic to our universal decoder. 

In the second part of this Supplementary Information, we have compared our approach with other pre-existing techniques. The key differences between the present scheme and previous proposals and experiments are: ({\it i}) applicability to any quantum communication task,  and not just useful for cryptography or non-locality; ({\it ii}) validity for arbitrary misalignments that occur on any time scale; and ({\it iii}) encoding of logical qubits in single photons, while all previous schemes not restricted to just cryptography or non-locality use at least a pair of photons per qubit.

The most significant feature of the presented hybrid-encoding scheme is that it does not restrict to non-locality or QKD. It allows in contrast for fully general misalignment-immune QC protocols, including for instance quantum teleportation, dense coding, or entanglement swapping, which is the basic component of quantum repeaters.